\documentclass[twocolumn,prd,superscriptaddress,preprintnumbers,nofootinbib]{revtex4}
\usepackage{graphicx}
\usepackage{epsfig}
\usepackage{lipsum}
\usepackage{enumitem}
\usepackage{bm}
\usepackage{latexsym,amssymb,amsmath,amsfonts,amssymb,txfonts,pxfonts,wasysym,float}
\usepackage{color}

\newcommand{\beq}[1]{\begin{equation}\label{#1}}
\newcommand{\eeq}{\end{equation}}
\newcommand{\bea}[1]{\begin{eqnarray} \label{#1}}
\newcommand{\eea}{\end{eqnarray}}
\newcommand{\ba}{\begin{array}}
\newcommand{\ea}{\end{array}}

\def\be{\begin{equation}}
\def\ee{\end{equation}}
\def\gs{\mathrel{
   \rlap{\raise 0.511ex \hbox{$>$}}{\lower 0.511ex \hbox{$\sim$}}}}
\def\ls{\mathrel{
   \rlap{\raise 0.511ex \hbox{$<$}}{\lower 0.511ex \hbox{$\sim$}}}}

\newcommand{\postscript}[2]{\setlength{\epsfxsize}{#2\hsize}
   \centerline{\epsfbox{#1}}}

\newcommand{\comment}[1]{}

\usepackage[usenames,dvipsnames]{xcolor}
\definecolor{orange}{cmyk}{0,0.5,1,0}
\definecolor{rossoCP3}{cmyk}{0,.88,.77,.40}
\definecolor{graa}{rgb}{0.8,0.8,0.8}
\definecolor{blaa}{rgb}{0.2,0.2,0.6}

\begin{document}

\title{\color{rossoCP3}{Through the Looking-Glass with ALICE into the
    Quark-Gluon Plasma:\\
 A New Test for Hadronic Interaction Models
    Used in Air Shower Simulations}}

\author{\bf Luis A. Anchordoqui}

\affiliation{Department of Physics and Astronomy,  Lehman College, City University of
  New York, NY 10468, USA
}

\affiliation{Department of Physics,
 Graduate Center, City University
  of New York,  NY 10016, USA
}

\affiliation{Department of Astrophysics,
 American Museum of Natural History, NY
 10024, USA
}

\author{\bf Carlos Garc\'{\i}a Canal}

\affiliation{Instituto de
  F\'{\i}sica La Plata, UNLP, CONICET Departamento de F\'{\i}sica,
  Facultad de Ciencias Exactas, Universidad Nacional de La Plata,
  C.C. 69, (1900) La Plata, Argentina}

\author{\bf Sergio J. Sciutto}

\affiliation{Instituto de
  F\'{\i}sica La Plata, UNLP, CONICET Departamento de F\'{\i}sica,
  Facultad de Ciencias Exactas, Universidad Nacional de La Plata,
  C.C. 69, (1900) La Plata, Argentina}

\author{\bf Jorge F. Soriano}

\affiliation{Department of Physics and Astronomy,  Lehman College, City University of
  New York, NY 10468, USA
}

\affiliation{Department of Physics,
 Graduate Center, City University
  of New York,  NY 10016, USA
}

\begin{abstract}
  \vskip 2mm \noindent Recently, the ALICE Collaboration reported an
  enhancement of the yield ratio of strange and multi-strange hadrons
  to charged pions as a function of multiplicity at mid-rapidity in
  proton-proton, proton-lead, lead-lead, and xenon-xenon
  scattering. ALICE observations provide a strong indication that a
  quark-gluon plasma is partly formed in high multiplicity events of
  both small and large colliding systems. Motivated by ALICE's
  results, we propose a new test for hadronic interaction models used
  for analyzing ultra-high-energy-cosmic-ray (UHECR) collisions with
  air nuclei. The test is grounded in the almost equal column-energy
  density in UHECR-air collisions and lead-lead collisions at the LHC.
  We applied the test to post-LHC event generators describing hadronic
  phenomena of UHECR scattering and show that these QCD Monte Carlo-based
  codes must be retuned to accommodate the strangeness enhancement
  relative to pions observed in LHC data.
  \end{abstract}

\maketitle

Besides addressing key questions in astrophysics, ultra-high-energy
cosmic ray (UHECR) experiments provide unique access to particle
physics at energies an order-of-magnitude higher center-of-mass energy than
$pp$ collisions at the Large Hadron Collider
(LHC)~\cite{Anchordoqui:2018qom}. However, a precise characterization
of the particle physics properties is usually hampered by the
ambiguity of model predictions computed through extrapolation of
hadronic interaction models tuned to accommodate collider data. These
predictions have sizable
differences~\cite{Anchordoqui:1998nq,Ulrich:2010rg,dEnterria:2011twh},
even among modern (post-LHC)
models~\cite{Calcagni:2017tws}, and quite often overlap with the phase of
particle physics observables. Disentangling one from the other is of
utmost importance to study particle physics in unexplored regions of
the phase-space. The development of new approaches to reduce the
systematic uncertainties of hadronic interaction models represents one
of the most compelling challenges in UHECR data analysis. In this
Letter we introduce a reliable technique for extrapolation into the ultra-high-energy domain.

QCD calculations on the Lattice~\cite{Borsanyi:2010cj} predict that under certain critical
conditions of baryon number density and temperature, normal nuclear
matter undergoes a phase transition to a deconfined state of quarks
and gluons where chiral
symmetry is restored~\cite{Shuryak:1980tp}. For many purposes, such a quark-gluon plasma
(QGP) can be described as a 
near-perfect fluid with surprisingly large
entropy-density-to-viscosity ratio. Therefore, once formed, like any
other hot object, the QGP transfers heat internally by radiation. Several phases can be identified during the QGP evolution. The
initial state contains only gluons as well as valence $u$ and $d$
quarks, but strangeness is produced in the very early stages via hard (perturbative) $2 \to 2$ partonic scattering
processes ($gg \to s\bar s$ and $q\bar q \to s \bar s$). Strangeness
is also predominantly produced during the subsequent partonic evolution via
gluon splittings ($g \to s\bar s$). This is because the very high
baryochemical potential inhibits gluons from fragmenting into $u\bar
u$ and $d\bar d$, and therefore they fragment predominantly into
$s\bar s$ pairs~\cite{Rafelski:1982pu}. In the hadronization process that follows this leads
to the strong suppression of pions (and hence photons), but allows the
production of heavy
hadrons with high transverse momentum ($p_T$) carrying away
strangeness.  At low $p_T$ non perturbative processes dominate the
production of strange hadrons. Thus, the abundances of strange particles
relative to pions provide a powerful
discriminator to  identify the  QGP formation.

A QGP can be created by heating nuclear matter up to a temperature of $2 \times 10^{12}~{\rm K}$, which amounts to 175~MeV per
particle. Relativistic heavy-ion collisions are then the best tool one
has to search for QGP production. Recently, the  ALICE Collaboration
reported enhancement of the yield ratio of multi-strange hadrons to
charged pions as a function of multiplicity at mid-rapidity in
LHC proton-proton ($pp$), proton-lead ($p$Pb), lead-lead (PbPb), and
xenon-xenon (XeXe) collisions~\cite{ALICE:2017jyt,Palni:2019ckt,Noferini:2019dzb,Vertesi:2019awk}. More concretely:
\begin{itemize}[noitemsep,topsep=0pt]
  \item the production rate of $K_S^0$, $\Lambda$, $\phi$, $\Xi$, and
    $\Omega$ increases with multiplicity faster than that for charged  particles;
\item the higher the strangeness content of the hadron, the more pronounced is the increase;
\item the ratios do not seem to depend on the system size or collision energies.
\end{itemize}
Altogether, this provides unambiguous evidence
for the formation of a QGP in high multiplicity small and large colliding systems~\cite{Koch:2017pda}.

Now, if the QGP is formed in relativistic heavy-ions collisions one
would also expect to be formed in the scattering of UHECRs in the
upper atmosphere~\cite{Farrar:2013sfa,Anchordoqui:2016oxy}. Moreover, since the column-energy density in
UHECR-air collisions is comparable to that in PbPb collisions at the
LHC, the precise characterization of the QGP properties from ALICE
data enables us to investigate QGP models describing the scattering of
cosmic rays that impinge on the Earth's atmosphere with energy
$10^{9} \alt E/{\rm GeV} \alt 10^{11}$. Indeed, as we show herein
ALICE data straightforwardly constrain these models without the need
to rely on energy extrapolation.

Before proceeding, we pause to note that the column-energy density is
the relevant parameter to compare QGP models with experimental
data. This is 
because in the center-of-mass the particles are extremely Lorentz
contracted so the time it takes to pass through each other is small
compared to the time for signals to propagate transversely, and hence the
pertinent parameter is the total {\it surface energy
  density}.  The best way of getting this point
across is to consider the collision of two nuclei of baryon number
$A_1$ and $A_2$ in the center-of-mass frame. The energies per nucleon
for each nucleus are written as $E_1 = \sqrt{s}/(2 A_1)$ and
$E_2 = \sqrt{s}/(2 A_2)$, where $s$ denotes the total center of mass
energy squared. Approximating each nucleus in its rest-frame as a cube
of side $L = A ^{1/3}$ gives the surface energy density in
GeV/nucleon-cross-section~\cite{Farrar:2019cid}
\begin{equation}
\Sigma = A_1^{1/3} E_1 + A_2^{1/3} E_2 = \frac{1}{2} \ \sqrt{s}
\left(A_1^{-2/3} + A_2^{-2/3} \right) \, .
\label{eq:xi}
\end{equation}
Finally, following the de-facto standard of high-energy physics, we rewrite (\ref{eq:xi}) in
the nucleon-nucleon center-of-mass frame
\begin{equation}
  \Sigma = \frac{1}{4} \ \sqrt{s_{NN}} \ \left(A_1^{-2/3} + A_2^{-2/3}
  \right) \  (A_1 + A_2) \, ,
\label{eq:xi2}
\end{equation}
where $\sqrt{s_{NN}} = 2 \sqrt{s}/(A_1 + A_2)$
is the
center-of-mass energy per nucleon.

For LHC PbPb scattering at
$\sqrt{s_{NN}} = 5.02~{\rm TeV}$  we can use (\ref{eq:xi2}) to obtain 
\begin{equation}
\Sigma_{\rm LHC}^{\rm PbPb}  = 2.9 \times 10^4~{\rm
  GeV} \, ,
\end{equation}
whereas for LHC XeXe scattering at $\sqrt{s_{NN} } = 5.44~{\rm TeV}$,
we have
\begin{equation}
\Sigma_{\rm LHC}^{\rm XeXe}  = 1.2 \times 10^4~{\rm
  GeV} \, .
\end{equation}
This must be compared to UHECR protons colliding with air
nuclei at $10^{10.5} \alt s/{\rm GeV}^2  \alt  10^{12.5}$, which leads to 
\begin{equation}
 9.8 \times 10^4  < \Sigma_{\rm UHECR}^{p {\rm air}}/{\rm GeV} < 9.8 \times 10^5 \,,
\end{equation}
where we have taken $A_{\rm air} = 14$. For the same primary energy, if the UHECR is a nucleus instead of proton the column energy
density is reduced. Now, using (\ref{eq:xi}) it is
straightforward to see that for helium and carbon nuclei with $E \agt 10^9~{\rm
  GeV}$, $\Sigma_{\rm UHECR}^{A{\rm  \, air}} > \Sigma_{\rm LHC}^{\rm PbPb} $,
but already for nitrogen (and of course nuclei with larger baryon number) there is a
particular energy where $\Sigma_{\rm UHECR}^{ A  \, {\rm air}} \simeq
\Sigma_{\rm LHC}^{\rm PbPb} $. For example, when a nitrogen with $E
\simeq 10^{9}~{\rm GeV}$ collides with an air nucleus, we have $\sqrt{s_{NN}}
\simeq 12~{\rm TeV}$ and a column-energy
density $\Sigma_{\rm UHECR}^{{\rm N \, air}} \simeq 2.9 \times 10^4~{\rm
  GeV}$, which is comparable to $\Sigma_{\rm LHC}^{\rm
  PbPb}$. Therefore, under the well justified assumptions of universality  between different
projectile/target combinations and approximate independence of
the collision energy, we conjecture that 
the QGP model predictions of these two scattering processes must be
roughly the same. In particular, both LHC PbPb scattering at
$\sqrt{s_{NN}} = 5.02~{\rm TeV}$ and UHECR nitrogen-air collisions at
$\sqrt{s_{NN}} \simeq 12~{\rm TeV}$ should produce the same hadron-to-pion
yield ratios as a function of the charged multiplicity. The hadron-to-pion yield ratios as a function of the charged
multiplicity observed in LHC PbPb scattering at
$\sqrt{s_{NN}} = 5.02~{\rm TeV}$ have been reported by the ALICE
Collaboration~\cite{ALICE:2017jyt,Palni:2019ckt,Noferini:2019dzb,Vertesi:2019awk}, providing a direct calibration for  hadronic interaction models
used for analyzing UHECR collisions with air nuclei.

\begin{table}\centering
\caption{Selected particle species $\alpha$. \label{tabla}}
  \begin{tabular}{cc}
\hline
~~~~~~~~~~~~~~~$\alpha$~~~~~~~~~~~~~~~&~~~~~~~~~~~~~~~particles~~~~~~~~~~~~~~~\\\hline\hline
$\pi$&$\pi^++\pi^-$\\
$p$&$p^++\bar p$\\
$K$&$K_S^0$\\
$\Lambda$&$\Lambda+\bar\Lambda$\\
$\Xi$&$\Xi^-+\bar \Xi^+$\\
$\Omega$&$\Omega^-+\bar\Omega^+$\\\hline
\hline
\end{tabular}
\end{table}

The column energy density is subject to large fluctuations from
collision to collision. For fixed nucleon-nucleon center-of-mass
energy, the multiplicity of charged secondary particles is expected to
be a reasonable tracer of the column energy density. Large
multiplicities correspond to many nucleons interacting (high density),
small multiplicities to few nucleons participating in the collision
(low density). Taking this argument into account one can perform a
comparison of prediction to data as a function of charged particle multiplicity instead of the
non-observable column energy density. Because charged multiplicity is
a good tracer of the energy density in the collision, the particle ratios
are expected to depend on whether the QGP is formed (or not) in the collisions. This
is very well seen in the ALICE data~\cite{ALICE:2017jyt,Palni:2019ckt,Noferini:2019dzb,Vertesi:2019awk}. High secondary multiplicities
correspond to the formation of a larger QGP region than
low-multiplicity interactions, as expected. Furthermore, the observed
particle ratios are, to a first approximation, only depending on the
charged particle multiplicity (in the considered energy range). They
are similar for a given charged particle multiplicity and independent
of the projectile-target combinations and different nucleon-nucleon
center-of-mass energies. This can then be interpreted as reflecting the conjectured dependence on the column energy density.

\begin{figure}[tpb] 
\begin{minipage}[t]{0.5\textwidth}
\postscript{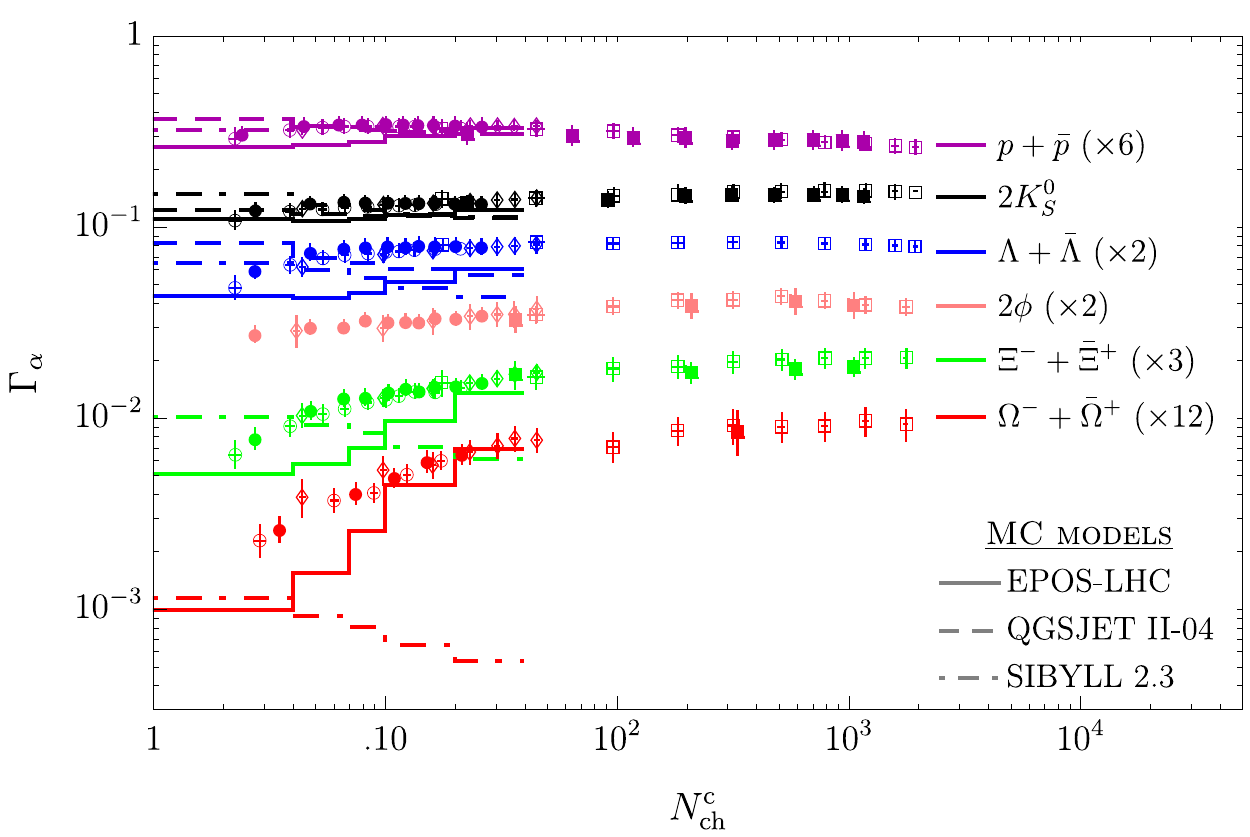}{0.9}
\end{minipage}
\begin{minipage}[t]{0.5\textwidth}
\postscript{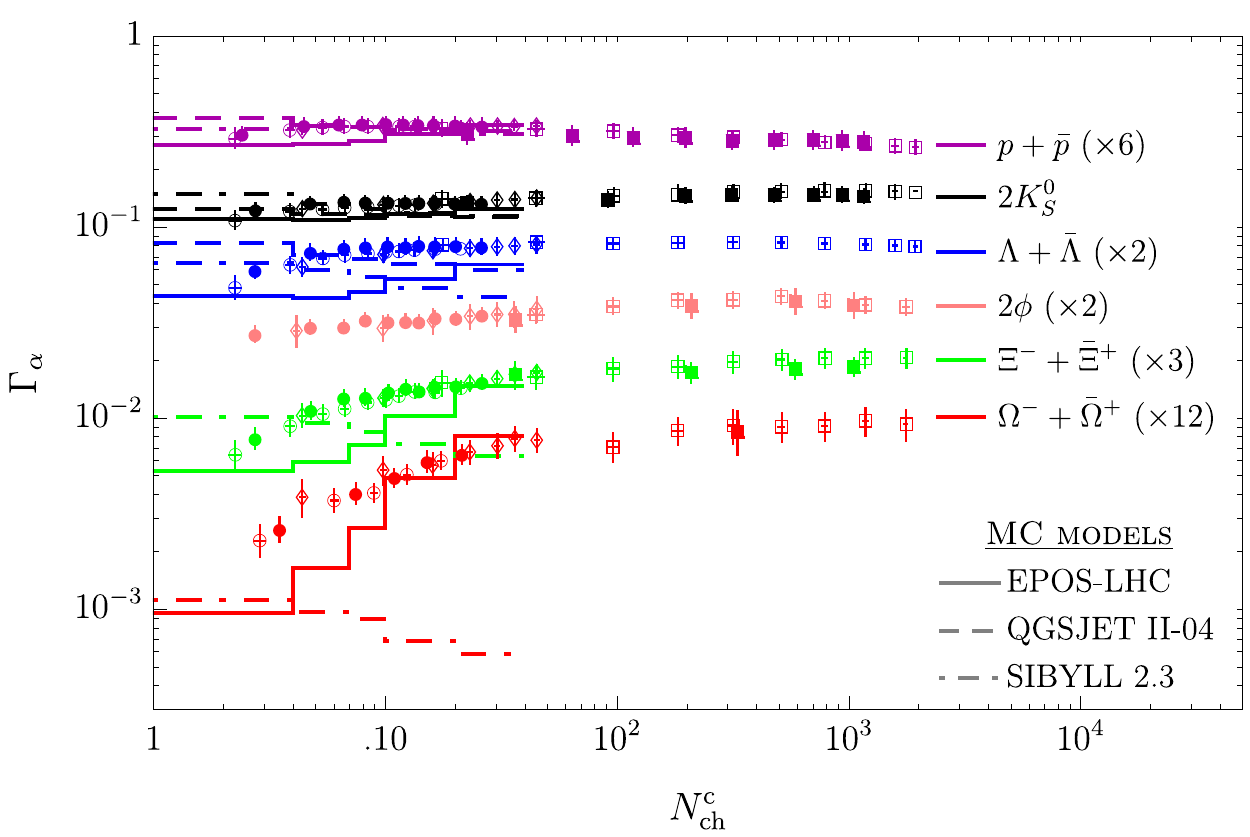}{0.9}
\end{minipage}
\begin{minipage}[t]{0.5\textwidth}
\postscript{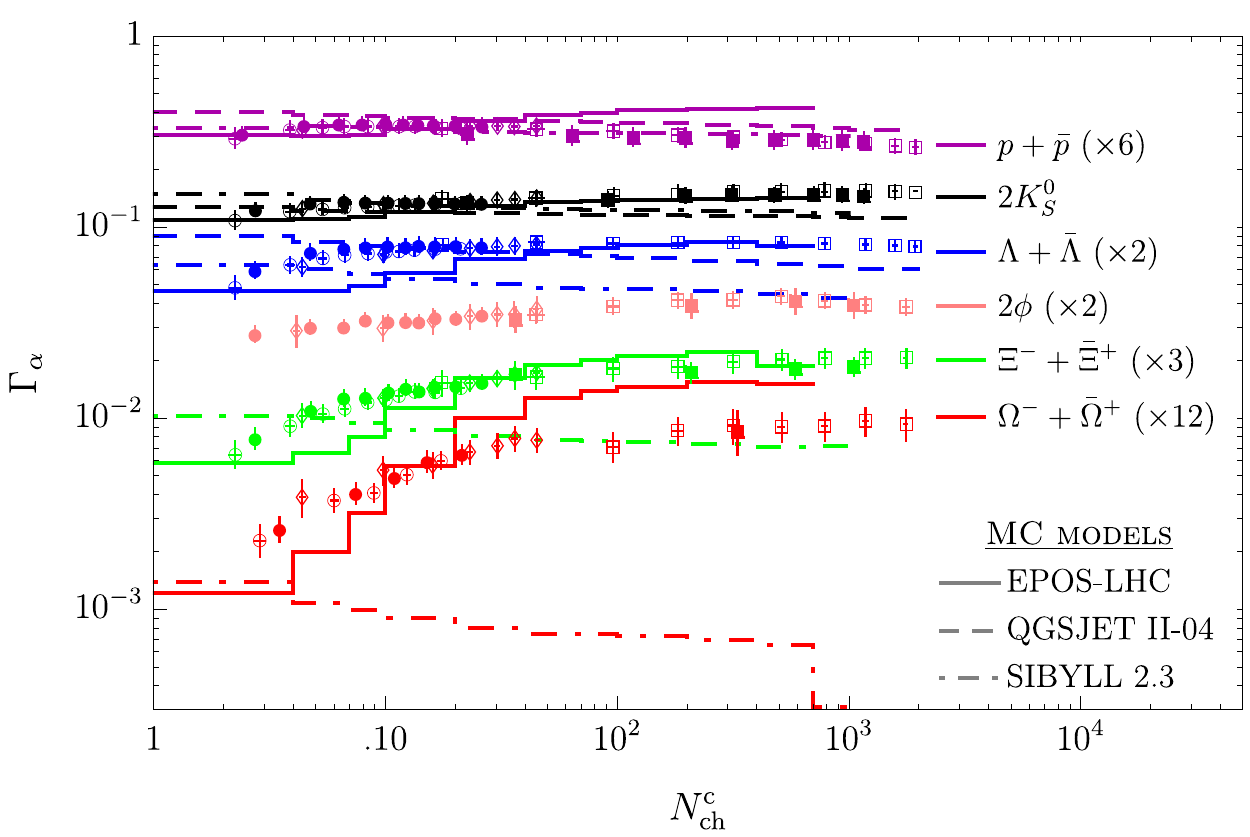}{0.9}
\end{minipage}
\caption{Hadron-to-pion yield ratios as a function of the charged
  particle multiplicity in $pp$, $p$Pb, PbPb, and XeXe collisions at
  the LHC. The predictions of post-LHC hadronic interaction models
  (top-to-bottom, $pp$ $\sqrt{s} = 7~{\rm TeV}$, $pp$ $\sqrt{s} = 13~{\rm
    TeV}$,  NN $\sqrt{s_{NN}} = 12~{\rm TeV}$)  are  compared to data reported by the ALICE Collaboration: $\circ$ $pp$
  at $\sqrt{s} = 7~{\rm TeV}$, $\bullet$ $pp$
  $\sqrt{s} = 13~{\rm TeV}$, $\diamond$ $p$Pb at
  $\sqrt{s_{NN}} = 5.02~{\rm TeV}$, $\Box$ PbPb at
  $\sqrt{s_{NN}} = 5.02~{\rm TeV}$, $\blacksquare$ XeXe at
  $\sqrt{s_{NN}} = 5.44~{\rm TeV}$~\cite{Palni:2019ckt}. (We
    have corrected a factor of two which is missing in 
 the labeling of  $\Gamma_{\Lambda \bar \Lambda}$ in Fig.~6
 of~\cite{Palni:2019ckt},  Fig.~4 of~\cite{Noferini:2019dzb}, 
 Fig.~1 of~\cite{Vertesi:2019awk}, and Fig.~1 of~\cite{Sharma:2019dne}.)  \label{fig:1}}
\end{figure}

We now turn to compare the predictions of post-LHC hadronic
interaction models (QGSJET II-04~\cite{Ostapchenko:2010vb} , EPOS-LHC~
\cite{Pierog:2013ria}, and SIBYLL 2.3c~\cite{Fedynitch:2018cbl,Engel:2019dsg}) with
the experimental data reported by the ALICE
Collaboration~\cite{Palni:2019ckt}. We run  $10^6$ collisions for each of the models, pair of primary particles, and center-of-mass energy. In analogy with the analyses presented by the ALICE Collaboration, we select those collisions containing at least one charged particle within the central ($|\eta|<1$) pseudorapidity region. For those collisions, we first select the charged particles at midrapidity ($|\eta|<0.5$). To estimate the observable $\left<dN_{\rm ch}/d\eta\right>_{|\eta|<0.5}$, we write it as 
\begin{eqnarray}
\left<dN_{\rm
    ch}/d\eta\right>_{|\eta|<0.5} & = &\frac{\int_{|\eta|<0.5}\dfrac{dN_{\rm
      ch}}{d\eta}d\eta}{\int_ {|\eta|<0.5}d\eta} =  N_{\rm
  ch}({|\eta|<0.5}) \nonumber \\
& \equiv & N_{\rm ch}^{\rm c} \, ,
\label{cinco}
\end{eqnarray}
the total number of charged particles at midrapidity which, for the
$i$-th collision, is denoted by $N_{\mathrm{ch},i}^{\rm c}$. For this collision, we measure the total number of particles $N_{\alpha,i}$ of several groups of species $\alpha$, as described in Table~\ref{tabla}.
Armed with (\ref{cinco}), we obtain the ratios to charged pions as
\begin{equation}
\Gamma_{\alpha,i}\equiv\frac{N_{\alpha,i}}{N_{\pi,i}}.
\end{equation}
In Fig.~\ref{fig:1} we show the average ratios
$\Gamma_\alpha\equiv \left<\Gamma_{\alpha,i}\right>$ to all the
collisions with the same $N_{\rm ch}^{\rm c}$ for the six species
listed in Table~\ref{tabla} as reported by the ALICE
Collaboration. For comparison, we also show the predictions of EPOS-LHC
and SIBYLL 2.3c for the above mentioned species (other than $\phi$)
considering $pp$ collisions $\sqrt{s} = 7~{\rm TeV}$ and
$\sqrt{s} = 13~{\rm TeV}$, as well as NN collisions at
$\sqrt{s_{NN}} = 12~{\rm TeV}$. We note, however, that the particles
that play a role on the evolution of UHECR showers are pions, kaons,
protons, neutrons, lambdas (and the corresponding antiparticles). For
the simulations run with QGSJET, we only display predictions for the
relevant secondaries driving the shower evolution. Overall, we
conclude that none of the models correctly reproduce the main
tendencies of ALICE data, especially for the description of
multi-strange hadron production. For $pp$ collisions, all hadronic
interaction models seem to reproduce quite well $\Gamma_{p \bar p}$ and
$\Gamma_{K_S^0}$, but fail to reproduce
$\Gamma_{\Lambda \bar \Lambda}$. For NN collisions, EPOS-LHC reaches a
good enough standard to pass the test in predicting the number of
secondary kaons and lambdas as a function of the charge
multiplicity. However, $\Gamma_{p\bar p}$ is overproduced by roughly
25\%. SIBYLL 2.3c provides a good description of $\Gamma_{p\bar p}$,
but fails to predict the number of kaons and lambdas. Finally, QGSJET
slightly overproduces $\Gamma_{p\bar p}$ and fails to predict
$\Gamma_{K_S^0}$ and $\Gamma_{\Lambda \bar \Lambda}$. All in all,
EPOS-LHC provides the best description of the hadron-to-pion yield
ratios as a function of the charged multiplicity relevant in the
modelling of UHECR shower evolution. Of course, if QGP effects are
correctly implemented in the models they should describe the
aforementioned features as seen in data.

We end with three observations:
\begin{itemize}[noitemsep,topsep=0pt]
\item Over the last year there has been a
tremendous amount of progress in modeling UHECR interactions with
EPOS-LHC~\cite{Pierog:2019opp}. In particular, the new EPOS-QGP has been properly tuned to
reproduce the particle to pion ratio for the $\Omega$ baryon versus
multiplicity at mid-rapidity as reported by the ALICE
Collaboration~\cite{Baur:2019cpv,Pierog:icrc}. It will be interesting to see whether the EPOS-QGP predictions of
NN collisions at $\sqrt{s_{NN}} = 12~{\rm TeV}$  can accurately match the experimental
data of $\Gamma_{p \bar p}$.
\item Future LHC data (including $p$O and OO
  collisions~\cite{Citron:2018lsq}) will provide new insights to guide software development.
\item The formation of a QGP could play a significant role in the
  development of UHECR air-showers. In particular, the enhanced
  production of multi-strange hadrons in high-multiplicity small and
  large colliding systems would suppress the fraction of energy which
  is transferred to the electromagnetic shower-component. The
  formation of QGP blobs in air showers would then enhance the number
  of muons reaching ground level, and would also modify the shape of the
  muon density distribution $\rho_\mu(r)$. The curvature of this
  distribution ($d^2\rho_\mu/dr^2$) has been proposed as a possible
  discriminator between hadronic interaction models with sufficient
  statistics~\cite{Anchordoqui:2003gm}. A thorough study of these
  phenomena is underway and will be presented
  elsewhere.
\end{itemize}

\section*{Acknowledgements}
We thank Prabi Palni and Francesco Noferini for valuable email communication. L.A.A. and
  J.F.S. are supported by U.S. National Science Foundation (NSF Grant
  PHY-1620661) and by the National Aeronautics and Space
  Administration (NASA 80NSSC18K0464). C.G.C. and S.J.S. are partially
  supported by ANPCyT.


\begin{thebibliography}{99}

\bibitem{Anchordoqui:2018qom} 
  L.~A.~Anchordoqui,
    {\color{rossoCP3} Ultra-high-energy cosmic rays},
  Phys.\ Rep.\  {\bf 801}, 1 (2019)
  doi:10.1016/j.physrep.2019.01.002
  [arXiv:1807.09645 [astro-ph.HE]].

\bibitem{Anchordoqui:1998nq} 
  L.~A.~Anchordoqui, M.~T.~Dova, L.~N.~Epele and S.~J.~Sciutto,
  {\color{rossoCP3}  Hadronic interactions models beyond collider energies},
  Phys.\ Rev.\ D {\bf 59}, 094003 (1999)
  doi:10.1103/PhysRevD.59.094003
  [hep-ph/9810384].

\bibitem{Ulrich:2010rg} 
  R.~Ulrich, R.~Engel and M.~Unger,
   {\color{rossoCP3} Hadronic multiparticle production at ultra-high energies and extensive air showers},
  Phys.\ Rev.\ D {\bf 83}, 054026 (2011)
  doi:10.1103/PhysRevD.83.054026
  [arXiv:1010.4310 [hep-ph]].
  
\bibitem{dEnterria:2011twh} 
  D.~d'Enterria, R.~Engel, T.~Pierog, S.~Ostapchenko and K.~Werner,
   {\color{rossoCP3}  Constraints from the first LHC data on hadronic event generators for ultra-high energy cosmic-ray physics},
  Astropart.\ Phys.\  {\bf 35}, 98 (2011)
  doi:10.1016/j.astropartphys.2011.05.002
  [arXiv:1101.5596 [astro-ph.HE]].


  
\bibitem{Calcagni:2017tws} 
  L.~Calcagni, C.~A.~Garc\'{\i}a Canal, S.~J.~Sciutto and T.~Tarutina,
    {\color{rossoCP3} LHC updated hadronic interaction packages analyzed up to cosmic-ray energies},
  Phys.\ Rev.\ D {\bf 98}, no. 8, 083003 (2018)
  doi:10.1103/PhysRevD.98.083003
  [arXiv:1711.04723 [hep-ph]].

\bibitem{Borsanyi:2010cj} 
  S.~Borsanyi, G.~Endrodi, Z.~Fodor, A.~Jakovac, S.~D.~Katz, S.~Krieg, C.~Ratti and K.~K.~Szabo,
    {\color{rossoCP3} The QCD equation of state with dynamical quarks},
  JHEP {\bf 1011}, 077 (2010)
  doi:10.1007/JHEP11(2010)077
  [arXiv:1007.2580 [hep-lat]].


\bibitem{Shuryak:1980tp} 
  E.~V.~Shuryak,
    {\color{rossoCP3} Quantum chromodynamics and the theory of superdense matter},
  Phys.\ Rept.\  {\bf 61}, 71 (1980).
  doi:10.1016/0370-1573(80)90105-2


    
\bibitem{Rafelski:1982pu} 
  J.~Rafelski and B.~Muller,
    {\color{rossoCP3} Strangeness production in the quark-gluon plasma},
  Phys.\ Rev.\ Lett.\  {\bf 48}, 1066 (1982)
  Erratum: [Phys.\ Rev.\ Lett.\  {\bf 56}, 2334 (1986)].
  doi:10.1103/PhysRevLett.48.1066, 10.1103/PhysRevLett.56.2334

\bibitem{ALICE:2017jyt} 
  J.~Adam {\it et al.} [ALICE Collaboration],
   {\color{rossoCP3} Enhanced production of multi-strange hadrons in high-multiplicity proton-proton collisions},
  Nature Phys.\  {\bf 13}, 535 (2017)
  doi:10.1038/nphys4111
  [arXiv:1606.07424 [nucl-ex]].
  


\bibitem{Palni:2019ckt} 
  P.~Palni (for the ALICE Collaboration),
  {\color{rossoCP3} Multiplicity dependence of strangeness and charged particle production in proton-proton collisions},
  arXiv:1904.00005 [nucl-ex].

 
\bibitem{Noferini:2019dzb} 
  F.~Noferini,
   {\color{rossoCP3} ALICE highlights},
  MDPI Proc.\  {\bf 13}, 6 (2019)
  doi:10.3390/proceedings2019013006
  [arXiv:1906.02460 [hep-ex]].

\bibitem{Vertesi:2019awk}
R.~Vertesi [ALICE Collaboration],
{\color{rossoCP3} Overview of recent ALICE results},
Contribution to: EDS Blois 2019 
[arXiv:1910.01981 [nucl-ex]].

\bibitem{Sharma:2019dne}
M.~Sharma [ALICE Collaboration],
  {\color{rossoCP3} Strangeness production in $p-$Pb collisions at 8.16~TeV},
[arXiv:1911.04845 [hep-ex]].


\bibitem{Koch:2017pda} 
  P.~Koch, B.~Müller and J.~Rafelski,
    {\color{rossoCP3} From strangeness enhancement to quark–gluon plasma discovery},
  Int.\ J.\ Mod.\ Phys.\ A {\bf 32}, no. 31, 1730024 (2017)
  doi:10.1142/S0217751X17300241
  [arXiv:1708.08115 [nucl-th]].

\bibitem{Farrar:2013sfa} 
  G.~R.~Farrar and J.~D.~Allen,
   {\color{rossoCP3} A new physical phenomenon in ultra-high energy collisions},
  EPJ Web Conf.\  {\bf 53}, 07007 (2013)
  doi:10.1051/epjconf/20135307007
  [arXiv:1307.2322 [hep-ph]].

  

\bibitem{Anchordoqui:2016oxy} 
  L.~A.~Anchordoqui, H.~Goldberg and T.~J.~Weiler,
     {\color{rossoCP3} Strange fireball as an explanation of the muon excess in Auger data},
  Phys.\ Rev.\ D {\bf 95}, no. 6, 063005 (2017)
  doi:10.1103/PhysRevD.95.063005
  [arXiv:1612.07328 [hep-ph]].

  
\bibitem{Farrar:2019cid} 
  G.~R.~Farrar,
    {\color{rossoCP3} Particle physics at ultrahigh energies},
  arXiv:1902.11271 [hep-ph].

\bibitem{Ostapchenko:2010vb} 
  S.~Ostapchenko,
   {\color{rossoCP3} Monte Carlo treatment of hadronic interactions in
     enhanced pomeron scheme I: QGSJET-II model},
  Phys.\ Rev.\ D {\bf 83}, 014018 (2011)
  doi:10.1103/PhysRevD.83.014018
  [arXiv:1010.1869 [hep-ph]].

 
\bibitem{Pierog:2013ria} 
  T.~Pierog, I.~Karpenko, J.~M.~Katzy, E.~Yatsenko and K.~Werner,
   {\color{rossoCP3} EPOS LHC: Test of collective hadronization with data measured at the CERN Large Hadron Collider},
  Phys.\ Rev.\ C {\bf 92}, no. 3, 034906 (2015)
  doi:10.1103/PhysRevC.92.034906
  [arXiv:1306.0121 [hep-ph]].

\bibitem{Fedynitch:2018cbl} 
  A.~Fedynitch, F.~Riehn, R.~Engel, T.~K.~Gaisser and T.~Stanev,
  {\color{rossoCP3} The hadronic interaction model Sibyll-2.3c and inclusive lepton fluxes},
  arXiv:1806.04140 [hep-ph].

\bibitem{Engel:2019dsg} 
  R.~Engel, A.~Fedynitch, T.~K.~Gaisser, F.~Riehn and T.~Stanev,
   {\color{rossoCP3} The hadronic interaction model Sibyll 2.3c and extensive air showers},
  arXiv:1912.03300 [hep-ph].


  


  
\bibitem{Pierog:2019opp} 
  T.~Pierog, B.~Guiot, I.~Karpenko, G.~Sophys, M.~Stefaniak and K.~Werner,
  {\color{rossoCP3} EPOS 3 and air showers},
  EPJ Web Conf.\  {\bf 210}, 02008 (2019).
  doi:10.1051/epjconf/201921002008

  
\bibitem{Baur:2019cpv} 
  S.~Baur, H.~Dembinski, T.~Pierog, R.~Ulrich and K.~Werner,
   {\color{rossoCP3} The ratio of electromagnetic to hadronic energy in high energy hadron collisions as a probe for collective effects, and implications for the muon production in cosmic ray air showers},
  arXiv:1902.09265 [hep-ph].

\bibitem{Pierog:icrc}
  T. Pierog, S. Baur, H. Dembinski, R. Ulrich, and K. Werner,
   {\color{rossoCP3} Collective hadronization and air showers: Can LHC
     data solve the muon puzzle?}
PoS ICRC {\bf 2019}, 387 (2019).


\bibitem{Citron:2018lsq} 
  Z.~Citron {\it et al.},
 {\color{rossoCP3}  Report from Working Group 5: Future physics opportunities for high-density QCD at the LHC with heavy-ion and proton beams},
  doi:10.23731/CYRM-2019-007.1159
  arXiv:1812.06772 [hep-ph].



\bibitem{Anchordoqui:2003gm} 
  L.~Anchordoqui and H.~Goldberg,
 {\color{rossoCP3}  Footprints of superGZK cosmic rays in the Pilliga State Forest},
  Phys.\ Lett.\ B {\bf 583}, 213 (2004)
  doi:10.1016/j.physletb.2003.12.072
  [hep-ph/0310054].
  


  
\end{thebibliography}
\end{document}